\documentclass[12pt]{article}
\usepackage{epsf}
\usepackage{amsfonts}
\usepackage{subfigure}

\def\df{{\rm d}}
\def\e{{\rm e}}
\def\GeV{{\rm GeV}}
\def\H{{\rm H}}

\newcommand{\be}{\begin{equation}}
\newcommand{\ee}{\end{equation}}
\newcommand{\bea}{\begin{eqnarray}}
\newcommand{\eea}{\end{eqnarray}}

\newcommand{\complex}{{{\rm I} \kern -.59em {\rm C}}}

\begin{document}
\begin{titlepage}
  \renewcommand{\thefootnote}{\fnsymbol{footnote}}
  \begin{flushright}
    \begin{tabular}{l@{}}
    astro-ph/0410133
    \end{tabular}
  \end{flushright}

  \vskip 0pt plus 0.4fill

  \begin{center}
    \textbf{\LARGE A model of inflation independent of
    the initial conditions,
    with bounded number of e-folds
    and $n_s$ larger or
    smaller than one}
  \end{center}

 \vskip 1ex plus 0.7fill

  \begin{center}
    {\large
    G.~Germ{\'a}n$^a$%
    \footnote{
    E-mail: gabriel@fis.unam.mx},
    A. de la Macorra$^b$%
    \footnote{E-mail: macorra@fisica.unam.mx}\\[0.3cm]
    }
    \textit{
  $^{(a)}$Centro de Ciencias F\'{\i}sicas, Universidad Nacional Aut{\'o}noma de M{\'e}xico,\\
  Apartado Postal 48-3, 62251 Cuernavaca, Morelos,  M{\'e}xico\\[0.3cm]
  $^{(b)}$Instituto de F\'{\i}sica, Universidad Nacional Aut{\'o}noma de
  M{\'e}xico,\\
  Apdo. Postal 20-364, 01000 M{\'e}xico D.F. M{\'e}xico}\\

    \vspace{1ex}
    \vskip 1ex plus 0.3fill

    {\large \today}

    \vskip 1ex plus 0.7fill

    \textbf{Abstract}
  \end{center}
  \begin{quotation}
We study a supergravity model of inflation essentially depending on one parameter
which can be identified with the slope of the potential at the origin. In this type
of models the inflaton rolls at high energy from negative values and a positive curvature
potential. At some point defined by the equation $\eta=1$ inflation starts. The potential
curvature eventually changes to negative values and inflation ends when $\eta=-1$.
No spontaneous symmetry breaking type mechanism for inflation of the new type to occur is
here required. The model naturally gives a bounded $\it{total}$ number of e-folds which is
typically close to the required number for observable inflation and it is independent of the
initial conditions for the inflaton. The energy scale introduced is fixed by the amplitude
of the anisotropies and is of the order of the supersymmetry breaking scale. The model can
also accommodate an spectral index bigger or smaller than one without extreme fine tuning.
We show that it is possible to obtain reasonable numbers for cosmological parameters and,
as an example, we reproduce values obtained recently by Tegmark $\it{et. al.}$, from WMAP
and SDSS data alone.

\end{quotation}

\vskip 0pt plus 2fill

\setcounter{footnote}{0}

\end{titlepage}

\section{Introduction}\label{intro}

We are certainly living very promising days for cosmology. In a
few years we have witnessed an explosion of cosmological data
recollection coming from telescope, balloon and satellite
experiments looking at the large scale structure distribution of
galaxies as well as cosmic microwave background radiation
anisotropies \cite {review1}. In particular projects like the
Wilkinson Microwave Anisotropy Probe WMAP \cite {wmap} and the
Sloan Digital Sky Survey SDSS \cite {sdss}, in spite of having yet
to complete their full schedule of observations, have already
provided a significant amount of data which has been used,
together with some reasonable assumptions, to determine
cosmological parameters in a variety of ways. In particular this
should eventually make possible to discriminate among several
competing models of inflation.

Here we study a supergravity model of inflation essentially
depending on one free parameter which can be identified with the slope
of the potential at the origin and an energy scale fixed by the amplitude
of the anisotropies, of the order of magnitude of the supersymmetry breaking
scale $10^{11} GeV$. The model naturally gives a
$\it{total}$ number of e-folds which is typically close to the
required number for observable inflation. The model can also
accommodate a spectral index bigger or smaller than one without
extreme fine tuning. We show that it is possible to obtain
reasonable numbers for cosmological parameters and, as an example,
we reproduce values obtained recently by Tegmark
$\it{et. al.}$ \cite {tegmark}, particularly the data provided in
the sixth column of Table 4 of that paper, where cosmological
parameters are determined from a six-parameter model using WMAP
and SDSS data alone. We choose this sample set of parameters for
simplicity.

The model described here actually defines a new type of models where the inflaton rolls at
high energy from negative values and a positive curvature potential. There is no inflation yet,
however at some point defined by the equation $\eta\equiv V''/V=1$ inflation starts. The potential
curvature eventually changes to negative values and inflation ends when $\eta=-1$, (see Fig.$1$).
No spontaneous symmetry breaking type mechanism for inflation of the new type to occur is here required.
As a consequence the $\it{total}$ number of e-folds
is bounded independently of the initial conditions for the inflaton and close to the required number
for observable inflation, see Table $1$. The spectral index can take values bigger than one at
the expense of increasing the slope of the potential and thus decreasing the total
number of e-folds.

In section $2$ we give a complete description of the model and obtain some general
expressions related to inflation in the slow-roll approximation. Section $3$ presents the main
results in the form of a table. We give a brief description of the ways the
essentially unique free parameter of the model is fixed and calculate all other relevant quantities.
Finally we conclude in section 4 briefly discussing the main results.

\section{The Model} \label{model}

We construct a model from the F-term part of the $N=1$ supergravity potential for a
single scalar field $z$. From now on we will take
$M\,(\equiv\,M_{\rm P}/\sqrt{8\pi}\simeq2.44\times10^{18}\GeV)=1$, where $M$ is the
normalized Planck mass. The supergravity potential is given by \cite {sugra}
\begin{equation}
 V = \e^K
     \left[F_z(K_{zz^*})^{-1}F_{z^*}^* -
     3 |W|^2\right] + {\rm D-terms} ,
\label{pot}
\end{equation}
where
\begin{equation}
 F_{z} \equiv \frac{\partial W}{\partial z} +
       \left(\frac{\partial K}{\partial z}\right) W ,\qquad
 K_{zz^*} \equiv \frac{\partial^2 K}{\partial z\partial z^*} ,
\end{equation}
and $K$ is the K\"{a}hler potential. To first approximation we take it to be of the
canonical form
\begin{equation}
 K(z,z^*)= z z^*+...,
\label{kp}
\end{equation}
and the superpotential $W(z)$ is determined imposing some reasonable physical
assumptions as follows \cite {w}.
The sectors of the theory in charge of supersymmetry breaking, inflation
and the visible sector interact among themselves only gravitationally. In particular,
this allow us to study the inflationary sector independently of the others.
Requiring that supersymmetry remain unbroken in the global minimum and that the
contribution to the cosmological constant vanishes we find that, at the minimum,
the superpotential and its first derivative should vanish. Thus a first contribution
to $W(z)$ in a Taylor expansion around the minimum at $z=z_0$ is
\begin{equation}
 W(z)  = f(z_0)(z-z_0)^2,
\label{sp}
\end{equation}
where $f(z_0)$ is a constant term with dimensions of mass. We take the simplest choice
\begin{equation}
f(z_0) = \Lambda.
\label{fun}
\end{equation}
We then write
\begin{equation}
z = \frac{1}{\sqrt{2}}(\phi+i \chi),
\label{zeta}
\end{equation}
and set $\chi=0$, being a stable direction of the full potential (\ref{pot})
 which is then given by
\begin{equation}
 V = \Lambda^2 \e^{\phi^2 /2}(\phi-\phi_0)^2(2+(\phi-\phi_0)(6\phi_0+\phi(2+\phi^2-\phi\phi_0))/8) .
\label{potencial}
\end{equation}
Finally we redefine the minimum at $\phi_0$ by
\begin{equation}
\phi_0 = \sqrt{2}+c/8,
\label{ficero}
\end{equation}
where $c$ is a constant which will be determined by requiring agreement with
cosmological parameters as given in \cite {tegmark}.  The parameter $c$ is essentially  the
only free parameter of the model, it dictates the $\it{shape}$ of the potential and thus the
dynamics of the inflaton while $\Lambda$ is just an
overall scale specifying the $\it{height}$ of the potential during inflation
$V \sim \Lambda^2 M^2$ which will be fixed by the amplitude of the anisotropies.
The choice Eq.(\ref{ficero})
makes easy to interpret $c$: it gives (to first order) the slope of the potential
at the origin since $V'(0)=c(1+3c\sqrt{2}/32+c^2/256)$.
Thus $c=0$ (the value used in all previous models \cite {ross-s}, \cite {w})
makes the potential flat at the origin. Here no extreme fine tuning in $c$ is required,
being ${\cal O}(10^{-4})$, with a modest tuning of the inflaton mass
$m_{\phi} \approx \sqrt{-c} \approx{\cal O}(10^{-2})$.
\begin{figure}[t]
\centering
\centerline{\epsffile{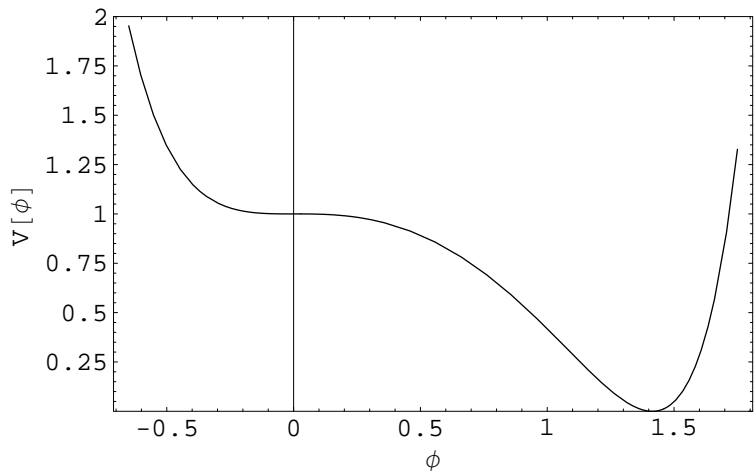}}
\caption{The inflationary potential (in units of $\Lambda^2 M^2$) Eq.(\ref{potencial})  for
$c=-1.18 \times 10^{-4}$ is shown as a function of $\phi$, the real part of the scalar field $z$,
Eq.(\ref {zeta}). For this value of $c$ the slope of the potential around the origin
is always negative,
being equal to $c$, to first order. Initially the would be inflaton rolls at high energy from
negative values without inflating.
When the condition $\eta = 1$ is reached inflation starts. We can easily
solve for the $\it{start}$
of inflation at $\phi_s \approx -0.094$. At some point the potential curvature changes sign from
positive to negative until inflation ends when $\eta = -1$ is satisfied.
In this example the end of
inflation occurs at $\phi_e \approx 0.169$ giving a $\it{total}$ number
of e-folds $N_T=137$, all
other relevant parameters are shown in Table 1. In this type of models we do not
assume a previous epoch where the inflaton is taken to the origin and kept there by thermal
effects waiting for spontaneous symmetry breaking to occur and inflation of the new type to start.
Rather the inflaton rolls from high energy to its global minimum with an inflationary
epoch in between. The modest tuning of $c$ is not a tuning on initial conditions (for large $|\phi|$
the potential is dominated by the exponential term alone and is insensitive to $c$.) but
a requirement for an intermediate stage of observable inflation to occur.}
\label{fig1}
\end{figure}
For negative $c$ the potential will always have negative slope
(except at the global minimum where, of course, it vanishes
becoming positive afterwards). This is shown in Fig.$1$. For small
positive $c$ the potential develops a local minimum and then a
maximum close to the origin. Because for negative $c$ the
potential slope is always negative we can precisely determine not
only the end but also the beginning of inflation, i.e., the total
number of e-folds is bounded. Thus we can also take the model as a
realization of a recent proposal by Banks and Fischler \cite {b-f}
where it is claimed that a universe which asymptotically becomes
de Sitter is consistent with inflationary models with a bound on
the total number of e-folds $N_{T}$ (see, however, \cite
{oponentes}). Our model has a bounded number of e-folds
independently of the initial conditions for the inflaton. The
supergravity potential shows that for large $|\phi|$-values
$V(\phi) \approx \e^{\phi^2}$. Thus the would be inflaton might
start rolling from anywhere at high energy without having any
effect on inflation. In fact the beginning of inflation starts at $\eta=1$
and expanding $\eta$ as a function of $\phi$ and $c$ and keeping
the lowest order one has $\eta=-3c/2\sqrt{2} - 6\sqrt{2}\phi+ O(c^2,\phi^2)\simeq 1$
giving a value $\phi_{s}=-1/6\sqrt{2}\simeq -0.1$. This
means that  no fine tuning on the value $\phi_i$
is needed and for any  $|\phi_i|$ larger than $|\phi_{s}|\simeq 0.1$
the number of efolds of inflation  will be the same (in general
one expects at high energies
an initial value $\phi_i$ of the order of the Planck mass).
The evolution
of $\phi$ is given by the usual equation of motion
\be
\ddot\phi+3H\dot\phi+V'=0
\ee
and since $V'$ is negative for $\phi<\phi_0$  we will have an increasing $\phi$
until it oscillates around $\phi_0$.
The inflationary stage is controlled by
the slope or shape parameter $c$ which affects the potential only
for small $\phi$ (see Fig.$1$). Characteristics of inflation such
as the number of e-folds, beginning and end of inflation, spectral
index and so on are controlled by the value of $c$ independently
of the initial conditions for the inflaton \cite {vacha}. We do
not assume a previous epoch where the inflaton is taken to the
origin and kept there by thermal effects waiting for spontaneous
symmetry breaking to occur and inflation of the new type to start.
Rather the inflaton rolls from high energy to its global minimum
with a transient inflationary epoch in between (see Fig.$1$). The
modest tuning of $c$ is not a tuning on initial conditions but a
requirement for an intermediate stage of observable inflation to
occur. We calculate the scalar spectral index with the usual
slow-roll expression \cite {review}
\begin{equation}
n_s = 1+2 \eta-6 \epsilon,
\label{indice}
\end{equation}
and its logarithmic derivative
\begin{equation}
\frac{d n_s}{d \ln k} = 16 \eta \epsilon-24 \epsilon^2-2 \xi,
\label{derivative}
\end{equation}
where $\eta$, $\epsilon$ and $\xi$ are the slow-roll parameters
\begin{equation}
\eta= \frac{V^{\prime\prime}}{V} ,
\label{eta}
\end{equation}
\begin{equation}
\epsilon= \frac{1}{2}\left(\frac{V^\prime}{V}\right)^2 ,
\label{epsilon}
\end{equation}
\begin{equation}
\xi= \frac{V^\prime V^{\prime\prime\prime}}{V^2} .
\label{xi}
\end{equation}
The amplitude of scalar fluctuations is given by \cite {peiris-verde}
\begin{equation}
A_s= \frac{1}{2.95 \times 10^{-9}}\frac{25}{4}\delta_\H^2,
\label{amplitud}
\end{equation}
where
\begin{equation}
\delta_\H^2 = \frac{1}{75 \pi^2}\frac{V^3}{{V^\prime}^2} =
\left(1.91 \times 10^{-5}\frac{\e^{1.01(1-n_s)}}{\sqrt{1+0.75 r}}\right)^2.
\label{delta}
\end{equation}
In the last equality of Eq.(\ref{delta})
we have used the fitting function of Bunn et. al., \cite {bunn}. The value of $\phi$ at
horizon crossing $\phi_*$
when $k_*=a H$ during inflation is given by assuming a value for the number of e-folds $N$
\bea
 N_{*} = -\int_{{\phi}_{*}}^{{\phi}_\e}
         \frac{V({\phi})}{V'({\phi})} \df{\phi}.
\label{number}
\eea
or solving consistently using some other criteria. In Table 1 we use various
criteria to determine $N_*$. In the expression above $\phi_e$ denotes the value of $\phi$
at the end of inflation, determined by the violation of the slow-roll $\eta=-1$.
The scale $\Lambda$ obtained from Eq.(\ref{delta}) turns out to be ${\cal O}(10^{11} GeV)$
and could be related to the supersymmetry breaking scale while the inflationary
scale $\Delta\equiv V^{1/4} \approx (\Lambda M)^{1/2}$ is of the order of $10^{14}-10^{15} GeV$.
The relative amplitude of the tensor to scalar modes $r$ is given by \cite {peiris-verde}
\begin{equation}
r = 16 \epsilon .
\label{ere}
\end{equation}
At the end of inflation the oscillations of the inflaton field would make it decay thus
reheating the universe. The couplings of the inflaton to  some other bosonic $\chi$ or
fermionic $\psi$ MSSM fields occur due to terms
$-\frac{1}{2}g^2\phi^2\chi^2$ or $-h\bar{\psi}\psi\phi$, respectively. These
couplings induce decay rates of the form \cite {linde}
\begin{equation}
\Gamma(\phi\to\chi\chi)=\frac{g^4\phi_0^2}{8\pi m_{\phi}},\qquad
\Gamma(\phi\to\bar{\psi}\psi)=\frac{h^2m_{\phi}}{8\pi}
\label{decays}
\end{equation}
where $\phi_0$ is the value of $\phi$ at the minimum of the potential Eq.(\ref {ficero})
and $m_{\phi}$ is the inflaton mass given by
\begin{equation}
m_{\phi} \approx \Lambda.
\label{inflamass}
\end{equation}
A maximum value for the decay is obtained when $m_{\chi,\psi}\approx
m_{\phi}$. In this case we find
\begin{equation}
\Gamma\approx \frac{m_{\phi}^3}{8\pi\phi_0^2}.
\label{decay}
\end{equation}
The reheat temperature at the beginning of the radiation-dominated era is thus
\cite{turner}
\begin{equation}
T_{rh} \approx \left( \frac{90}{\pi^2g_{*}}\right)^{\frac{1}{4}}\sqrt{\Gamma}
\approx \frac{1}{\sqrt{16 \pi}}\left( \frac{90}{\pi^2g_{*}}\right)^{\frac{1}{4}} \Lambda^{3/2},
\label{trh}
\end{equation}
where $g_*$ is the number of relativistic degrees of freedom which for
the MSSM equals $915/4$.

\section{Results}\label{results}

Our main results are presented in Table $1$ where we show, in particular, that the scalar spectral
index can have values bigger or smaller than one. We can understand this as follows. For small $c$
the slow-roll parameter $\epsilon$ is much smaller than $\eta$ during inflation thus we have
\begin{equation}
n_s \approx 1+2 \eta .
\label{indiceapp}
\end{equation}
Thus we can have $n_s > 1$ or $n_s < 1$ depending on the curvature of the potential. It is easy to
show that
\begin{equation}
\eta_* \approx -\frac{3}{2 \sqrt{2}}c-6\sqrt{2}\phi_*+\frac{39}{2}\phi^2_*+....
\label{etaapp}
\end{equation}
For negative $c$ the sign of $\eta_*$ depends on the relative size between the first and second
terms of Eq.(\ref {etaapp}). If we increase the slope of the potential by increasing the (absolute)
value of $c$ (but still ${\cal O}(10^{-4})$) $\phi_*$ takes smaller and eventually negative values to
satisfy the condition on the number of observable e-folds of inflation. At $\phi_* \approx -c/8$
curvature changes sign from negative to positive values and the spectral index can become bigger
than one at $\phi_*$. This actually defines a new type of models where the inflaton rolls at
high energy from negative values and a positive curvature potential without inflating due to the
dominance of the $\e^{\phi^2}$ term in the potential. For small $|\phi|$-values, at some point
defined by the equation $\eta=1$, inflation starts. The potential curvature eventually changes to
negative values and inflation ends when $\eta=-1$. No spontaneous symmetry breaking type mechanism
for inflation of the new type to occur is here required. We illustrate this behavior in Fig.$1$
and the rows of Table $1$ where $c$ takes negative values. Examples of $n_s > 1$ are shown in the second
and third rows where the spectral index takes the values $1.10$ and $1.00$
respectively. In both cases Eqs.(\ref{number}) and (\ref{liddle}) were solved consistently.
We next show that it is possible to obtain reasonable numbers for cosmological parameters.
As an example the first row shows cosmological parameters
relevant to inflation. These were obtained from a 6-parameter model by Tegmark $\it{et. al.}$,
(sixth column of Table 4 of \cite {tegmark}) using WMAP + SDSS data alone. We have taken that
particular column for simplicity. In the fourth row $(c=-1.18 \times 10^{-4})$ we determine
$c$ such that at $\phi_*$ we get the central value of the scalar spectral index such that the
number of e-folds Eq.(\ref {number}) be consistent with the recent upper estimate of Liddle and Leach
\cite {liddle} and Dodelson and Hui \cite {dodelson} for the maximum number of e-folds
for the observable universe.
\begin{equation}
N_* = 63.3 + \frac{1}{4}\ln{\epsilon}.
\label{liddle}
\end{equation}
Notice that this is not a bound on the $\it{total}$ number of e-folds which, in general, could be
much higher.
In the fifth row we simply set $N_{T}=50$ at $\phi_*$ and the sixth row requires a
$\it{total}$ number of e-folds to be equal to the upper estimate of Banks and
Fischler \cite {b-f}, $N_{T}=85$ giving $N_*=39$ at $\phi_*$. Finally the last
two rows show results for $c=0$ (flat case \cite {ross-s}) and a positive $c$ where
a local maximum develops close to the origin. In these two rows we do not impose
that the spectral index be $0.977$ at $\phi_*$ because that requirement violates the
bound Eq.(\ref {liddle}). Here we use Eq.(\ref {liddle}) consistently with Eq.(\ref {number})
to determine $\phi_*$ and then calculate all other quantities at $\phi_*$.

\begin{table}[tbp]
\begin{tabular}{|c|c|c|c|c|c|c|c|c|c|}
\hline
{\bf c} & ${\bf n_s}$ & ${\bf \frac{d n_s}{d \ln k}}$ & ${\bf A_s}$ & ${\bf r}$ & ${\bf
  \Delta}$ & ${\bf T_{rh}}$ & ${\bf N_*-N_{T}}$   \\
\hline
WMAP+SDSS & $0.977^{+0.039}_{-0.025}$ & 0 & $0.81^{+0.15}_{-0.09}$ & 0 & - & - & -   \\
-3.10 10$^{-4}$ & 1.10 & -0.008 & 0.63 & 2 10$^{-6}$ & 1 10$^{15}$ & 9 10$^{7}$ & 59-83   \\
-1.65 10$^{-4}$ & 1.00 & -0.003 & 0.77 & 2 10$^{-7}$ & 7 10$^{14}$ & 2 10$^{7}$ & 59-115   \\
-1.18 10$^{-4}$ & 0.977 & -0.002 & 0.81 & 1 10$^{-7}$ & 6 10$^{14}$ & 2 10$^{7}$ & 59-137   \\
-1.72 10$^{-4}$ & 0.977 & -0.003 & 0.81 & 3 10$^{-7}$ & 8 10$^{14}$ & 3 10$^{7}$ & 50-113   \\
-2.96 10$^{-4}$ & 0.977 & -0.005 & 0.81 & 7 10$^{-7}$ & 1 10$^{15}$ & 6 10$^{7}$ & 39-85    \\
0 & 0.93 & -0.001 & 0.89 & 3 10$^{-8}$ & 5 10$^{14}$ & 7 10$^{6}$ & 58-   \\
+1.18 10$^{-4}$ & 0.90 & -0.0007 & 0.94 & 1 10$^{-8}$ & 4 10$^{14}$ & 3 10$^{6}$ & 58-   \\
\hline
\end{tabular}
\caption{The first row shows cosmological parameters
relevant to inflation obtained from a 6-parameter model by Tegmark $\it{et. al.}$,
(sixth column of Table 4 of \cite {tegmark}) using WMAP + SDSS data alone. We have taken this
set of values for simplicity. The next two rows show the scalar
spectral index equal to $1.10$ and $1.00$ respectively. In both cases the number of e-folds
was required to be consistent with the recent upper estimate \cite {liddle}, \cite {dodelson}, given
by Eq.(\ref{liddle}) for the maximum number of e-folds for the observable universe.
Notice that this is not a bound on the $\it{total}$ number of e-folds which, in general, could
be much higher, in both cases we found $\phi_* < 0 $ and a positive curvature potential.
In the next three rows we determine $c$ such that at $\phi_*$ we get the central value of the
scalar spectral index $0.977$ and such that (fourth row) the number of e-folds Eq.(\ref {number})
be consistent with Eq.(\ref{liddle}).
In the fifth row we simply set $N_*=50$ at $\phi_*$ and the sixth row requires a
$\it{total}$ number of e-folds to be equal to the upper estimate of Banks and
Fischler \cite {b-f}, $N_{T}=85$ giving $N_*=39$ at $\phi_*$.  Finally the last
two rows show results for $c=0$ (flat case \cite {w}, \cite {ross-s}) and a positive $c$ where
a local maximum develops close to the origin. In these last two rows we do not impose
that the spectral index be $0.977$ at $\phi_*$ because that requirement violates the
bound Eq.(\ref {liddle}). We only use Eq.(\ref {liddle}) consistently with Eq.(\ref {number})
to determine $\phi_*$ and then calculate all other quantities at $\phi_*$. In all
cases we see that the running of the spectral index is very small compared with the present
sensitivity of observations. The scale of inflation $\Delta\equiv V(\phi_*)^{1/4}
\approx (\Lambda M)^{1/2}$ and the reheat temperature $T_{rh}$ in columns sixth and seventh
are given in $GeV$.
\label{table:1}}
\end{table}

\section{Conclusions}\label{con}

We have studied a very simple model of inflation which essentially
depends on one free parameter denoted by $c$ which controls the
shape of the potential for small $\phi$ and thus the
dynamics of inflation. The large-$|\phi|$ behavior, being dictated
by the exponential term of the potential, is insensitive to $c$. In this
sense characteristics of inflation are independent of the initial
conditions for the inflaton.  The parameter $c$ can be
identified with the slope of the potential at the origin. For
$c=0$ we have the flat case \cite {w}, \cite {ross-s} where the
slope and the curvature both vanish at $\phi=0$. In this case the
$\it{total}$ number of e-folds grows without limit as we approach
the origin and the spectral index can only take values lower than
one. For negative $c$, however, the slope of the potential during
inflation never vanishes and the amount of inflation is bounded,
i.e., the total number of e-folds is bounded. Finite negative $c$ values
actually define a new type of models where the would be inflaton rolls at
high energy from negative values and a positive curvature
potential without inflating. At some point defined by the equation $\eta=1$
inflation starts. The potential curvature eventually changes to negative
values and inflation ends when $\eta=-1$ (see Fig.$1$). Here the inflaton
rolls all the way from high energy down to its global minimum. No
pre-inflationary epoch where the inflaton is kept waiting at the origin
for spontaneous symmetry breaking to occur is here required.
The spectral index can take values even bigger than one at the expense
of increasing the slope of the potential and thus decreasing the
$\it{total}$ number of e-folds. We illustrate this in the second and
third rows of Table 1. We also show that it is possible to obtain
reasonable numbers for cosmological parameters and, as an example,
we reproduce values obtained recently by Tegmark
$\it{et. al.}$ \cite {tegmark}, where cosmological
parameters are determined from a six-parameter model using WMAP
and SDSS data alone. We see that the results
are in excellent agreement. Also, the reheat temperature is
sufficiently low to avoid overproduction of unwanted states such
as gravitinos.
The $\Lambda$ scale introduced in the superpotential can be
related to supersymmetry breaking in the hidden sector being ${\cal O}(10^{11} GeV)$
and the resulting scale of inflation $\Delta\equiv V(\phi_*)^{1/4}
\approx (\Lambda M)^{1/2}$ is close to the unification scale.
In conclusion the good agreement with the data in such a simple
model indicate that perhaps some of the underlying assumptions should
be part of a more elaborated model coming from a more fundamental theory.

\section{Acknowledgements}

G.G. would like to thank Graham G. Ross for discussions. This work
was supported by the project PAPIIT IN114903-3 and CONACYT:42096, 45718.

\end{document}